\documentclass[aps,preprint]{revtex4}%
\usepackage{amsfonts}
\usepackage{amsmath}
\usepackage{amssymb}
\usepackage{graphicx}%
\begin{document}
\title[Aharonov-Bohm phase shift]{Comment on Experiments Related to\\the Aharonov-Bohm Phase Shift}
\author{Timothy H. Boyer}
\affiliation{Department of Physics, City College of the City University of New York, New
York, New York 10031}

\begin{abstract}
Recent experiments undertaken by Caprez, Barwick, and Batelaan should clarify
the connections between classical and quantum theories in connection with the
Aharonov-Bohm phase shift. \ It is pointed out that resistive aspects for the
solenoid current carriers play a role in the classical but not the quantum
analysis for the phase shift. \ The observed absence of a classical lag effect
for a macroscopic solenoid does not yet rule out the possibility of a lag
explanation of the observed phase shift for a microscopic solenoid. \ 

\end{abstract}
\maketitle

\subsection{Introduction}

For the past thirty-five years, it has been suggested repeatedly that the
experimentally observed Aharonov-Bohm phase shift\cite{AB}\cite{C}\cite{MB}
may arise from a classical electromagnetic lag effect involving velocity
changes for charged particles passing a solenoid.\cite{B73b}\cite{B87}%
\cite{AC}\cite{B00a}\cite{B00b}\cite{B02a}\cite{B02b}\cite{B06a}\cite{B06b}
\ In recent years, these suggestions have been dismissed as being without
merit by the editors at the leading physics journals. \ The suggestions are
dismissed despite the fact that these semiclassical energy arguments predict
precisely the observed phase shift and despite the fact that the classical
electromagnetic interaction of a point charge and a solenoid remains poorly understood.

Within the past year, some of the ideas associated with the proposed classical
lag effect have come under experimental test from work by Caprez, Barwick, and
Batelaan. \ These experimenters observe no lag effect for electrons passing a
macroscopic solenoid.\cite{Bate} \ Of course, the Aharonov-Bohm phase shift
has never been observed for such a macroscopic solenoid, so this experiment
does not yet rule out a classical lag effect as the basis for the
Aharonov-Bohm phase shift. \ 

Indeed, the experiment with a macroscopic solenoid emphasizes the complete
contrast between the theoretical treatment of the Aharonov-Bohm phase shift
now appearing in all the recent quantum textbooks\cite{Sak} and the classical
electromagnetic analysis suggesting a lag effect. \ According to presently
accepted quantum theory, the Aharonov-Bohm phase shift arises from an enclosed
magnetic flux with no need to discuss any interaction between the passing
charges and the sources of the magnetic flux. \ In contrast, the classical lag
analysis depends crucially upon the details of the interaction between the
passing charged particles and the current-carriers of the magnetic
flux.\cite{B06a} \ 

One may ask how two such totally different points of view can lead to the same
experimental prediction. \ The relative success of the contrasting points of
view may reflect a situation analogous to that when a charged particle passes
a conductor. \ As long as the conductor is a good conductor, the forces on a
passing charged particle due to image charges will involve only the physical
shape of the conductor and not the detailed composition of the conductor.
\ However, if the conductivity is imperfect or if the induced charges
correspond to dielectric behavior, then more information regarding the
interaction of the passing charge and the material will be required to account
for the behavior of a passing charged particle. \ In a similar vein for the
Aharonov-Bohm phase shift, it may well be that when the resistive energy loss
of a solenoid is small, then all solenoids behave in the same way regarding
energy conservation for a passing charged particle. \ However, when the
resistive energy loss is large, then the energy-conserving interaction becomes
negligible and the interaction of the particle and solenoid becomes quite different.

\subsection{Summary of the Classical Aspects of the Aharonov-Bohm Situation}

The magnetic energy of interaction of a charged particle $q$ passing a
solenoid with constant currents is given by\cite{B1973}%

\begin{equation}
U=\frac{q}{c}\mathbf{v}_{q}\cdot\mathbf{A(r}_{q})
\end{equation}
where $\mathbf{v}_{q}$ is the velocity of the passing charge and
$\mathbf{A(r}_{q})$ is the vector potential in the Coulomb gauge of the
solenoid evaluated at the position of the passing charge. \ If this magnetic
interaction energy is compensated by a change in the kinetic energy of the
passing charge $mv_{q}\Delta v_{q}=-U$, then there is a relative spatial
lag\cite{B87}%
\begin{equation}
\Delta Y=\frac{q\Phi}{mv_{q}c}%
\end{equation}
between particles passing on opposite sides of the long solenoid of flux
$\Phi$, and the associated semiclassical phase shift is exactly that predicted
by Aharonov and Bohm\cite{AB}%
\begin{equation}
\Delta\phi=\frac{mv_{q}\Delta Y}{\hbar}=\frac{q\Phi}{c\hbar}%
\end{equation}
Associated with this spatial lag is also a relative time lag%
\begin{equation}
\Delta t=\frac{\Delta Y}{v_{q}}=\frac{q\Phi}{mv_{q}^{2}c}%
\end{equation}
between particles passing on opposite sides of the solenoid. \ However, the
currently accepted dogma is that the Aharonov-Bohm phase shift occurs without
any velocity changes for the passing charged particles. \ The experiments
recently undertaken by Herman Batelaan\cite{Batep} may allow one to
distinguish between these two competing points of view regarding the
experimentally observed phase shift.\ The classical electromagnetic
description based upon kinetic energy changes predicts the time delay in Eq.
(4) for charges passing on opposite sides of the solenoid. \ In the first, and
easiest experiment, Caprez, Barwick and Batelaan, looked for a time delay for
electrons passing a macroscopic solenoid. No delay was found although the
experimental accuracy was quite sufficient to detect a delay of the magnitude
in Eq. (4) predicted by energy conservation.\cite{Bate} \ 

If one accepts the current quantum point of view that all solenoids should
behave the same way regarding passing charged particles, then this experiment
already implies the failure of the classical lag explanation. \ However,
classical physics indicates that not all solenoids will interact in the same
way with passing charges. \ The relative magnitude of frictional (resistive)
effects plays an important role in classical descriptions of multiparticle systems.

\subsection{Analogue Involving Electromagnetic Induction}

In order for the observed Aharonov-Bohm phase shift to be based upon a change
in kinetic energy for a passing charge, we must have an electric field back at
the passing charge generated by induced currents in the solenoid which are
caused by the electric field of the passing charged particle.\cite{B02a}
\ Since a long constant-current solenoid has no electric or magnetic fields
outside its winding, any such back electric field must arise from
accelerations of the solenoid charges in response to the electric field of the
passing charge. \ Exactly such behavior has been calculated for the
interaction of a passing charge and a classical hydrogen atom\cite{B06a} where
the electromagnetic fields correspond to those of the Darwin
Lagrangian\cite{Jack}%
\begin{equation}
\mathbf{E=}\widehat{r}\frac{q}{r^{2}}\left(  1+\frac{1}{2}\frac{v^{2}}{c^{2}%
}-\frac{3}{2}\frac{(\mathbf{v\cdot}\widehat{r})^{2}}{c^{2}}\right)  -\frac
{q}{2c^{2}r}\left(  \mathbf{a}+\mathbf{a\cdot}\widehat{r}\widehat{r}\right)
\end{equation}%
\begin{equation}
\mathbf{B=}q\frac{\mathbf{v}}{c}\times\frac{\widehat{r}}{r^{2}}%
\end{equation}
giving the interaction of charged particles accurately through order $1/c^{2}$
when radiation is neglected. \ However, there has been no classical
calculation for a charged particle passing a multiparticle solenoid. \ 

We may come closest to understanding what is involved for a solenoid by
examining (as an analogy) a situation involving electromagnetic induction
where the accelerations of charged particles are again the crucial element.
\ Suppose we had a magnetic dipole $\overrightarrow{\mathbf{\mu}}%
\mathbf{=}\widehat{k}\mu$ oriented along the z-axis and moving with initial
velocity $\mathbf{v=}\widehat{k}v$ along the z-axis. \ The magnetic dipole
moves toward a set of $N$ equally-spaced charged particles in the plane z=0,
each of charge $e$ and mass $m,$ which are constrained to move on a circular
loop of radius $r$ with center on the z-axis. \ The moving magnetic moment
$\overrightarrow{\mathbf{\mu}}$ generates an electric field acting on each
charge $e$ given by%
\begin{equation}
E_{\phi}\mathbf{=-[v}_{\mu}\mathbf{\times B}_{\mu}\mathbf{/}c]_{\phi
}\mathbf{\ =}-3\frac{v_{\mu}}{c}\frac{rz}{\left(  z^{2}+r^{2}\right)  ^{5/2}%
}\mu
\end{equation}
The nonrelativistic equation of motion for each charge $e$ (neglecting
radiation) is%
\begin{equation}
mr\frac{d^{2}\phi}{dt^{2}}=-m\gamma r\frac{d\phi}{dt}+%
%TCIMACRO{\dsum \limits_{j=2}^{j=N}}%
%BeginExpansion
{\displaystyle\sum\limits_{j=2}^{j=N}}
%EndExpansion
\frac{-e^{2}}{2c^{2}}\frac{d^{2}\phi}{dt^{2}}\left(  2\sin\left\vert \frac{\pi
j}{N}\right\vert \right)  ^{-1}\left[  \cos\left(  \frac{2\pi j}{N}\right)
+\sin^{2}%
%TCIMACRO{\QOVERD{(}{)}{2\pi j}{N}}%
%BeginExpansion
\genfrac{(}{)}{}{}{2\pi j}{N}%
%EndExpansion
\right]  +eE_{\phi}%
\end{equation}
where $mr(d^{2}\phi/dt^{2})$ is the particle mass times acceleration,
$-m\gamma r(d\phi/dt)$ represents the frictional force on the particle due to
the constraining ring, the term with the sum involves the electric
acceleration fields in Eq. (5) of all the other charges on the ring acting on
the charge whose acceleration is being discussed, and $eE_{\phi}$ is the
electric force in Eq. (7) due to the approaching magnetic dipole. \ The
motions of the charges $e$ around the ring in turn cause a magnetic field
$\mathbf{B}_{e},$ which along the z-axis is given by%
\begin{equation}
\mathbf{B}_{e}=%
%TCIMACRO{\dsum }%
%BeginExpansion
{\displaystyle\sum}
%EndExpansion
\frac{e}{c}\frac{\mathbf{v}_{e}\times\widehat{r}}{r^{2}}=\widehat{k}N\frac
{e}{c}\frac{r^{2}}{(z^{2}+r^{2})^{3/2}}\frac{d\phi}{dt}%
\end{equation}
which acts back on the approaching magnetic dipole $\overrightarrow
{\mathbf{\mu}}$ with a force%
\begin{equation}
F_{x}=\frac{\partial}{\partial z}(\overrightarrow{\mathbf{\mu}}\mathbf{\cdot
B}_{e})=-3\mu N\frac{e}{c}\frac{zr^{2}}{(z^{2}+r^{2})^{5/2}}\frac{d\phi}{dt}%
\end{equation}
Here we have a clear physical situation involving accelerations of the
particles on the ring which in turn put a force back on the approaching
magnetic moment. \ The situation has an analogy to the classical interaction
of a point charge and a solenoid; both situations depend on the electric field
associated with the moving object to produce accelerations of the charges of a
multiparticle system, and then the force back on the moving object arises due
to changes in the multiparticle motions. \ 

The energies for the situation of Eqs. (7)-(10) involve the kinetic energy of
the magnetic dipole $\overrightarrow{\mathbf{\mu}}$\textbf{, }the kinetic
energy of the ring particles $e$, the magnetic energy of the ring particles,
and the magnetic interaction energy between the magnetic field of the ring
particles and the magnetic field of the magnetic dipole $\overrightarrow
{\mathbf{\mu}}\mathbf{.}$ If the particles $e$ on the ring are initially at
rest, then any energy transferred to these particles must come from the
approaching magnetic dipole. \ If the magnetic dipole passes through the
center of the circular ring of particles, then we expect that the magnetic
moment will be slowed down during its passage and hence that there will be a
relative time delay associate with the interaction of the particles on the
ring compared to a magnetic dipole $\overrightarrow{\mathbf{\mu}}$ which did
not encounter the ring.

We notice immediately that the acceleration $d^{2}\phi/dt^{2}$ in Eq. (8)
appears as a multiplicative factor both for the particle mass $m$ and for sum
term arising from the electric fields due to the accelerations of the other
charges $e.$ \ Thus if there are many charges so that this sum term is
enormous, then one can neglect the particle masses $m$ compared to the
self-inductance effects associated with the mutual interactions of the ring
charges $e$. \ Similarly, the particle kinetic energy and the magnetic energy
of self-inductance both depend upon the square of the velocities $r^{2}%
(d\phi/dt)^{2}$of the charges $e.$ \ We can see that we expect different
behavior when the friction constant $\gamma$ is large compared to when it is
small. \ If the friction constant $\gamma$ is very small, then we have a
situation involving energy conservation during the interaction of the magnetic
moment and the particles $e$ of the ring. \ We expect that the the particles
$e$ of the ring will accelerate in response to the electric field $E_{\phi}$
of the approaching magnetic dipole $\overrightarrow{\mathbf{\mu}},$ so that
kinetic energy will be removed from the magnetic dipole on its approach, and
then, after the magnetic dipole passes through the ring, the kinetic energy of
the magnetic dipole will be restored. \ The interaction of the magnetic dipole
and the ring will be evident in a relative time lag compared to a magnetic
dipole which continued with constant velocity. \ On the other hand, if the
friction constant $\gamma$ is very large, then from Eq. (8) we see that the
velocities $rd\phi/dt$ of particles on the ring will be very small since the
velocities depend inversely on $\gamma$. \ But the force back on the
approaching magnetic dipole depends upon the velocity $rd\phi/dt$ of the ring
particles, and hence the force back on the approaching magnetic moment will be
negligible for large $\gamma$. \ Hence no time lag for the passing magnetic
dipole will be observed for a large friction constant $\gamma.$ \ For a single
ring particle ($N=1),$ the relative size of the friction requires a comparison
between the value of $\gamma$ and the inverse passage time $v_{\mu}/r$ of the
magnetic moment through the circular ring. \ If the passage time is small
(large inverse passage time $\gamma<<v_{\mu}/r),$ then we expect $mr(d^{2}%
\phi/dt^{2})$ to dominate $mr\gamma(d\phi/dt)$. \ When many charges $e$ are
present on the ring ($N>>1),$ then the energy conservation requires that the
characteristic time $L/R$ of the ring as an $LR$ circuit should be long
compared to the passage time.

The same qualitative features arising in this example of a magnetic dipole
interacting with the charges on a ring will also occur for the interaction of
a charge particle passing a solenoid. \ In this case the accelerations of the
solenoid particles are not symmetrical and are causes by the \textit{electric}
field of the passing charge. \ When the frictional forces are very small, we
expect to find a time lag associated with the energy-conserving interaction
between the charged particle and the solenoid. \ On the other hand, if the
frictional forces (solenoid resistance) is large, then we do not expect to
find the time lag because the accelerations of the solenoid charges will be
small and hence the back forces at the passing charge will be small.

In the experiments of Moellenstedt and Bayh,\cite{MB} where the Aharonov-Bohm
phase shift is clearly present, the passage times of electrons past the
solenoids is of the order of $10^{-13}$ sec (for a 40 keV electron passing 20
microns from the center of the solenoid). This time is not much longer than
the collision time $10^{-14}$ sec in the Drude model for conductivity of a
metal. Indeed, Jackson\cite{Jack312} gives $\gamma$ as of the order of
$10^{13}$ inverse seconds where $-m\gamma v$ is the resistive damping of a
particle of mass $m$ and speed $v$. Thus it is possible that the conservation
of energy involving magnetic fields holds in the short-time regime where
Moellenstedt and Bayh's experiments were performed, yet would not hold for the
much longer passage times for the slower electrons passing the much larger
solenoid in the experiment of Caprez, Barwick, and Batelaan.

\subsection{Crucial Experimental Tests}

The crucial test involving time delay corresponds (in a single experiment) to
observing the Aharonov-Bohm phase shift and yet not observing the time delay
of Eq. (4). Such an observation would rule out the classical lag
interpretation of the Aharonov-Bohm phase shift. For the regimes where the
Aharonov-Bohm phase shift has been observed, the time delays would be
extraordinarily small. Thus for the experiments of Moellenstedt and Bayh, the
time delay in Eq. (4) is of the order of $10^{-21}$ sec.\ 

It is the shifts of particle interference patterns which allow the measurement
of extraordinarily small quantities. \ Thus use of the shift of the
interference pattern by a large flux seems the best way realistically to test
for the absence of a velocity change. \ According to the currently accepted
quantum point of view, the Aharonov-Bohm phase shift is a purely topological
shift, and there is no velocity change and hence no spatial lag for charged
particles passing on opposite sides of a solenoid. \ Accordingly, one can
increase the flux in a solenoid by an arbitrary amount and never break down
the particle interference pattern. \ On the other hand, in the classical lag
point of view, a sufficiently large solenoid flux will cause a relative lag
which is larger than the coherence length associated with the charged particle
and at this point the observed particle interference pattern should break
down. \ Caprez, Barwick, and Batelaan are hoping to achieve appropriate
conditions to observe this possible breakdown.

In the classical analysis, the observed Aharonov-Bohm phase shift due to a
solenoid (a line of \textit{magnetic} dipoles) is analogous to the observed
Mattucci-Pozzi phase shift\cite{MP} due to a line of \textit{electric}
dipoles.\cite{B02b} \ This is not at all the view of currently accepted
quantum theory, which accepts the idea that the electrostatic phase shift is
due to a classical lag but claims that the magnetic phase shift is due to a
new quantum mechanical effect having no classical analogue. \ Thus ideally one
would like to see both these phase shifts (electric and magnetic) tested at
the same time with the same coherence length for the electron beam. From both
classical and quantum analyses of the electrostatic situation, we believe that
we know exactly what is going on involving electrostatic forces leading to a
relative lag effect producing the Mattucci-Pozzi electrostatic phase shift. If
the interference pattern breaks down for large \textit{electric} dipole
magnitude but not for large \textit{magnetic} dipole magnitude, then (in
contradiction to the classical analysis) the mechanisms for these two
interference pattern shifts are quite different, as is indeed claimed by
currently accepted quantum theory. However, if the interference pattern breaks
down for both the electric and the magnetic phase shifts, the currently
accepted quantum view is in error.

\end{document}